\documentclass[a4paper,twoside,12pt]{article}
\input{obsjkt.sty}

\newcommand{\Msun}{\ensuremath{~{\rm M}_\odot}}                   
\newcommand{\Rsun}{\ensuremath{~{\rm R}_\odot}}                   
\newcommand{\rhosun}{\ensuremath{~\rho_\odot}}                    
\newcommand{\Teff}{\ensuremath{T_{\rm eff}}}                      
\newcommand{\TeffA}{\ensuremath{T_{\rm eff,A}}}                   
\newcommand{\TeffB}{\ensuremath{T_{\rm eff,B}}}                   
\newcommand{\logg}{\ensuremath{\log g}}                           
\newcommand{\Vsys}{\ensuremath{V_\gamma}}                         
\newcommand{\EBV}{\ensuremath{E(B\!-\!V)}}                        
\newcommand{\degr}{\ensuremath{^\circ}}                           
\renewcommand{\kms}{~km~s$^{-1}$}                                 
\newcommand{\chir}{\ensuremath{\chi_\nu^{\,2}}}                   

\newcommand{\etal}{\textit{et al.}}                               

\newcommand{\kepler}{\textit{Kepler}}
\newcommand{\tess}{\textit{TESS}}
\newcommand{\hip}{\textit{Hipparcos}}
\newcommand{\gaia}{\textit{Gaia}}

\newcommand{\targ}{ZZ~UMa}
\newcommand{\targfull}{ZZ Ursae Majoris}

\newcommand{\Msunnom}{\hbox{$\mathcal{M}^{\rm N}_\odot$}}
\newcommand{\Rsunnom}{\hbox{$\mathcal{R}^{\rm N}_\odot$}}
\newcommand{\Lsunnom}{\hbox{$\mathcal{L}^{\rm N}_\odot$}}

\usepackage{rotating}

\begin{document} 

\OBSheader{Rediscussion of eclipsing binaries: \targ}{J.\ Southworth}{2022 December}

\OBStitle{Rediscussion of eclipsing binaries. Paper XI. \\ ZZ Ursae Majoris, a solar-type system showing total eclipses and a radius discrepancy}

\OBSauth{John Southworth}

\OBSinstone{Astrophysics Group, Keele University, Staffordshire, ST5 5BG, UK}

\OBSabstract{\targ\ is a detached eclipsing binary with an orbital period of 2.299~d that shows total eclipses and starspot activity. We used five sectors of light curves from the Transiting Exoplanet Survey Satellite (\tess) and two published sets of radial velocities to establish the properties of the system to high precision. The primary star has a mass of $1.135 \pm 0.009$\Msun\ and a radius of $1.437 \pm 0.007$\Rsun, whilst the secondary component has a mass of $0.965 \pm 0.005$\Msun\ and a radius of $1.075 \pm 0.005$\Rsun. The properties of the primary star agree with theoretical predictions for a slightly super-solar metallicity and an age of 5.5~Gyr. The properties of the secondary star disagree with these and all other model predictions: whilst the luminosity is in good agreement with models the radius is too large and the temperature is too low. These are the defining characteristics of the \emph{radius discrepancy} which has been known for 40 years but remains an active area of research. Starspot activity is evident in the out-of-eclipse portions of the light curve, in systematic changes in the eclipse depths, and in emission at the Ca H and K lines in a medium-resolution spectrum we have obtained of the system. Over the course of the \tess\ observations the light and surface brightness ratios between the stars change linearly by 20\% and 14\%, respectively, but the geometric parameters do not. Studies of objects showing spot activity should account for this by using observations over long time periods where possible, and by concentrating on totally-eclipsing systems whose light curves allow more robust measurements of the physical properties of the system.}


\section*{Introduction}

Detached eclipsing binaries (dEBs) are our primary source of measurements of the physical properties of normal stars \cite{Torres++10aarv,Me15aspc} and have many astrophysical applications \cite{Torres++10aarv,Andersen91aarv,Me21univ}. dEBs containing stars of similar mass to our Sun are useful in particular in helping to constrain our understanding of stellar theory in the mass regime around the solar fiducial point, for example the amounts of internal mixing and convective core overshooting \cite{Spada+13apj,KirkbyKent+16aa,Graczyk+16aa}. Solar-type dEBs can also be helpful in studying starspot configurations \cite{Jeffers+06mn,Wang+21mn,Wang+22mn} and the radius discrepancy whereby low-mass stars are found to be systematically larger and cooler than predicted by theoretical models \cite{Spada+13apj,Torres13an,MorrellNaylor19mn}.

In this work we determine the physical properties of the dEB \targfull\ based on published radial velocity (RV) measurements and a light curve recently obtained by the Transiting Exoplanet Survey Satellite (\tess). Basic information on \targ\ is given in Table~\ref{tab:info}. The $B$ and $V$ magnitudes come from the Tycho star mapper \cite {Hog+00aa} on the \hip\ satellite \cite{Hipparcos97}, and are the average of 213 measurements well distributed in orbital phase; they agree well with the dedicated observations by Lacy \cite{Lacy92aj}. The 2MASS $JHK$ magnitudes \cite{Cutri+03book} are single-epoch and were obtained at orbital phase 0.223.


\targ\ was discovered to be a dEB by Kippenhahn \cite{Geyer++55kvb,JaniashviliLavrov89ibvs}. Photometry and light curve solutions have been reported by multiple authors \cite{JaniashviliLavrov89ibvs,Doppner62,LavrovLavrova88atsir,Clement+93iauc,Clement+97aas,Clement+97aas2}. Lacy \cite{Lacy92aj,Lacy02aj} measured its $V$ magnitude, $B-V$ and $U-B$ colour indices, and photometric indices in the $uvby\beta$ system.

\begin{table}[t]
\caption{\em Basic information on \targ. \label{tab:info}}
\centering
\begin{tabular}{lll}
{\em Property}                            & {\em Value}                 & {\em Reference}                   \\[3pt]
\textit{Hipparcos} designation            & HIP 51411                   & \cite{Hipparcos97}                \\
\textit{Tycho} designation                & TYC 4144-400-1              & \cite{Hog+00aa}                   \\
\textit{Gaia} EDR3 designation            & 1049087857023628672         & \cite{Gaia21aa}                   \\
\textit{Gaia} EDR3 parallax               & $5.5494 \pm 0.0140$ mas     & \cite{Gaia21aa}                   \\          
\tess\ Input Catalog designation          & TIC 138505004               & \cite{Stassun+19aj}               \\
$B$ magnitude                             & $10.43 \pm 0.03$            & \cite{Hog+00aa}                   \\          
$V$ magnitude                             & $9.83  \pm 0.02$            & \cite{Hog+00aa}                   \\          
$J$ magnitude                             & $8.713 \pm 0.029$           & \cite{Cutri+03book}               \\
$H$ magnitude                             & $8.412 \pm 0.021$           & \cite{Cutri+03book}               \\
$K_s$ magnitude                           & $8.334 \pm 0.014$           & \cite{Cutri+03book}               \\
Spectral type                             & G0 V + G8 V                 & \cite{Clement+97aas}              \\[3pt]
\end{tabular}
\end{table}

Popper \cite{Popper96apjs} reported obtaining 13 high-resolution spectra which showed double lines, as part of a survey of 76 late-type dEBs. He presented a short discussion and preliminary properties, but no RVs or detailed analysis. Popper \cite{Popper95ibvs} gave minimum masses of $M_1\sin^3i = 1.18$\Msun\ and $M_2\sin^3i = 0.96$\Msun, where $i$ is the orbital inclination, but again no RVs were presented in his short summary paper.

Lacy \& Sabby \cite{LacySabby99ibvs} obtained and measured RVs from 27 high-resolution \'echelle spectra of \targ. They combined these with the photometric results from Clement \etal\ \cite{Clement+97aas2} to obtain the first determination of the masses and radii of the component stars. Imbert \cite{Imbert02aa} presented a spectroscopic orbit of \targ\ based on 49 spectra from the \textit{Coravel} and \textit{\'Elodie} instruments. The resulting masses and radii are in reasonable agreement with those from Lacy \& Sabby.

In this work we revisit \targ\ to determine its physical properties to high precision. We base our analysis on the RVs from Lacy \& Sabby \cite{LacySabby99ibvs} and Imbert \cite{Imbert02aa}, and on light curves from \tess. A detailed scientific motivation is presented in Paper~I of this series \cite{Me20obs} and a review of the use of space-based photometry for the study of binary systems is given in Ref.\ \cite{Me21univ}.


\section*{Observational material}

\begin{figure}[t] \centering \includegraphics[width=\textwidth]{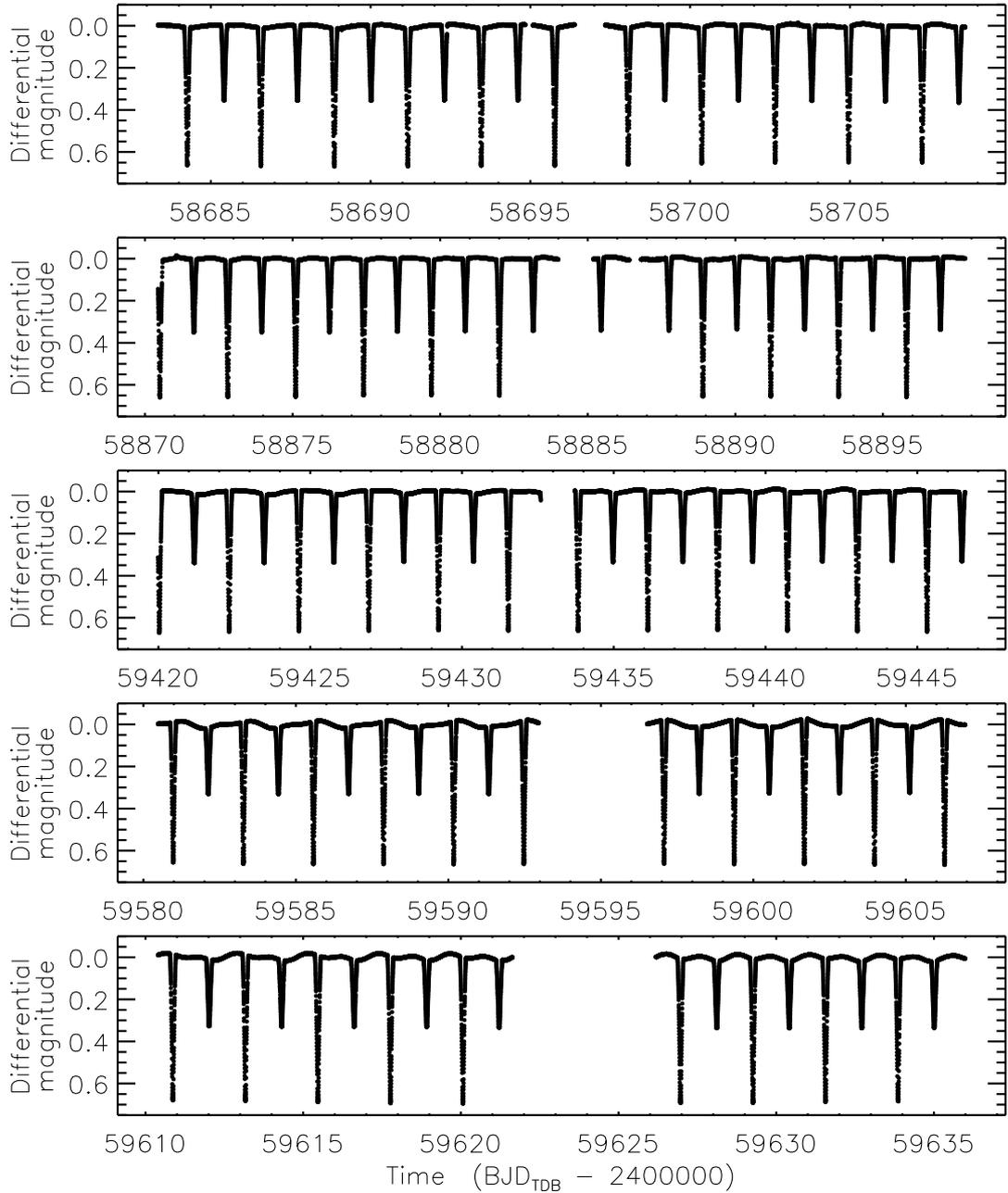} \\
\caption{\label{fig:time} \tess\ short-cadence SAP photometry of \targ\ from the
five sectors. The flux measurements have been converted to magnitude units then
rectified to zero magnitude by the subtraction of low-order polynomials.} \end{figure}


A surfeit of photometry exists for \targ\ from the NASA \tess\ satellite \cite{Ricker+15jatis}, which observed it in short cadence (120~s sampling rate) in sectors 14 (2019/07/18 to 2019/08/15), 21 (2020/01/21 to 2020/02/18), 41 (2021/07/23 to 2021/08/20) and 47--48 (2021/12/03 to 2022/02/26). The light curves show deep total and annular eclipses plus a smaller-amplitude and longer-timescale variation which changes between and during sectors and can be attributed to starspots.

We downloaded the data for all five sectors from the MAST archive\footnote{Mikulski Archive for Space Telescopes, \\ \texttt{https://mast.stsci.edu/portal/Mashup/Clients/Mast/Portal.html}} and converted the fluxes to relative magnitude. We retained observations with a QUALITY flag of zero, yielding a total of 84\,764 datapoints. We found the simple aperture photometry (SAP) and pre-search data conditioning SAP (PDCSAP) data \cite{Jenkins+16spie} to be visually almost indistinguishable, so adopted the SAP data as usual in this series of papers. The light curves are shown in Fig.~\ref{fig:time}.


\begin{figure}[t] \centering \includegraphics[width=\textwidth]{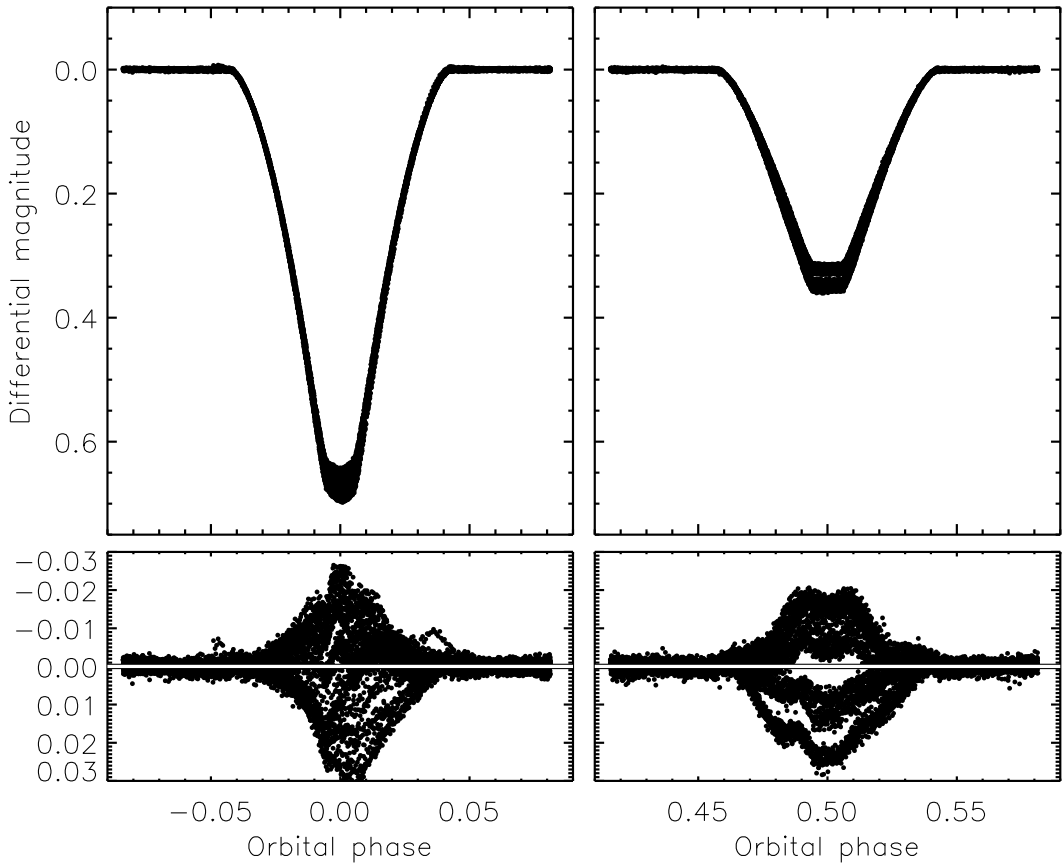} \\
\caption{\label{fig:phase} Best fit to the full \tess\ light curve of \targ\ using {\sc jktebop}
for the primary eclipse (left panels) and the seconday eclipse (right panels).
The residuals are shown on an enlarged scale in the lower panels.} \end{figure}

\section*{Light curve analysis}

The light curve shows total eclipses plus a slower variation of lower amplitude due to starspots. Evolution of the starspot pattern is clear both between and during sectors. A detailed analysis of the spot properties of \targ\ is not the aim of the current work; we instead view it as a nuisance signal to be removed\footnote{A model of the starspots and their evolution could be obtained using computer codes such as {\sc macula} (Kipping \cite{Kipping12mn}) or {\sc soap-T} (Oshagh \etal\ \cite{Oshagh+13aa2}).}. To this end we located every fully-observed eclipse within the \tess\ light curve and extracted all data within one eclipse duration of the eclipse centre. A straight line was fitted to the data outside eclipse and subtracted from the light curve to normalise it to zero differential magnitude. The result of this was a new dataset containing 26\,682 datapoints (31.5\% of the original number).

\begin{figure}[t] \centering \includegraphics[width=\textwidth]{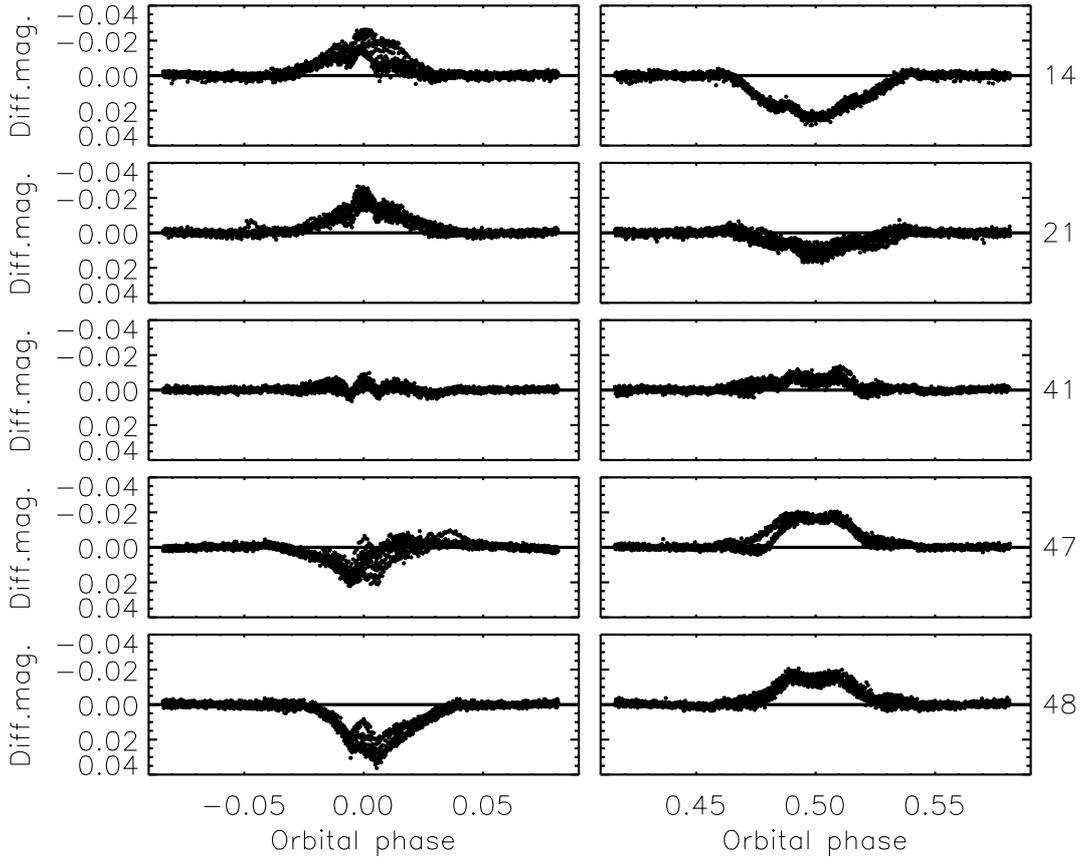} \\
\caption{\label{fig:resid1} Residuals of the best fit shown in Fig.~\ref{fig:phase}, but
separated according to \tess\ sector. The sector numbers are annotated on the right.} \end{figure}

We then fitted this eclipse light curve using version 42 of the {\sc jktebop}\footnote{\texttt{http://www.astro.keele.ac.uk/jkt/codes/jktebop.html}} code \cite{Me++04mn2,Me13aa}. We fitted for the orbital period ($P$) and time of mid-eclipse ($T_0$), the sum ($r_{\rm A}+r_{\rm B}$) and ratio ($k = {r_{\rm B}}/{r_{\rm A}}$) of the fractional radii, the orbital inclination ($i$), and the central surface brightness ratio of the two stars ($J$). We adopted a quadratic limb darkening (LD) law, fitted for the linear coefficients for each star ($u_{\rm A}$ and $u_{\rm B}$) and fixed the quadratic coefficients ($v_{\rm A}$ and $v_{\rm B}$) to theoeretical values from Claret \cite{Claret17aa}. A circular orbit was adopted as we found the orbital eccentricity to be extremely small and consistent with zero. Third light was found to be insignificant and typically slightly below zero so was fixed at zero. We term the star eclipsed at the deeper eclipse star~A and its companion star~B; star~A is hotter, larger and more massive than star~B.

\begin{figure}[t] \centering \includegraphics[width=\textwidth]{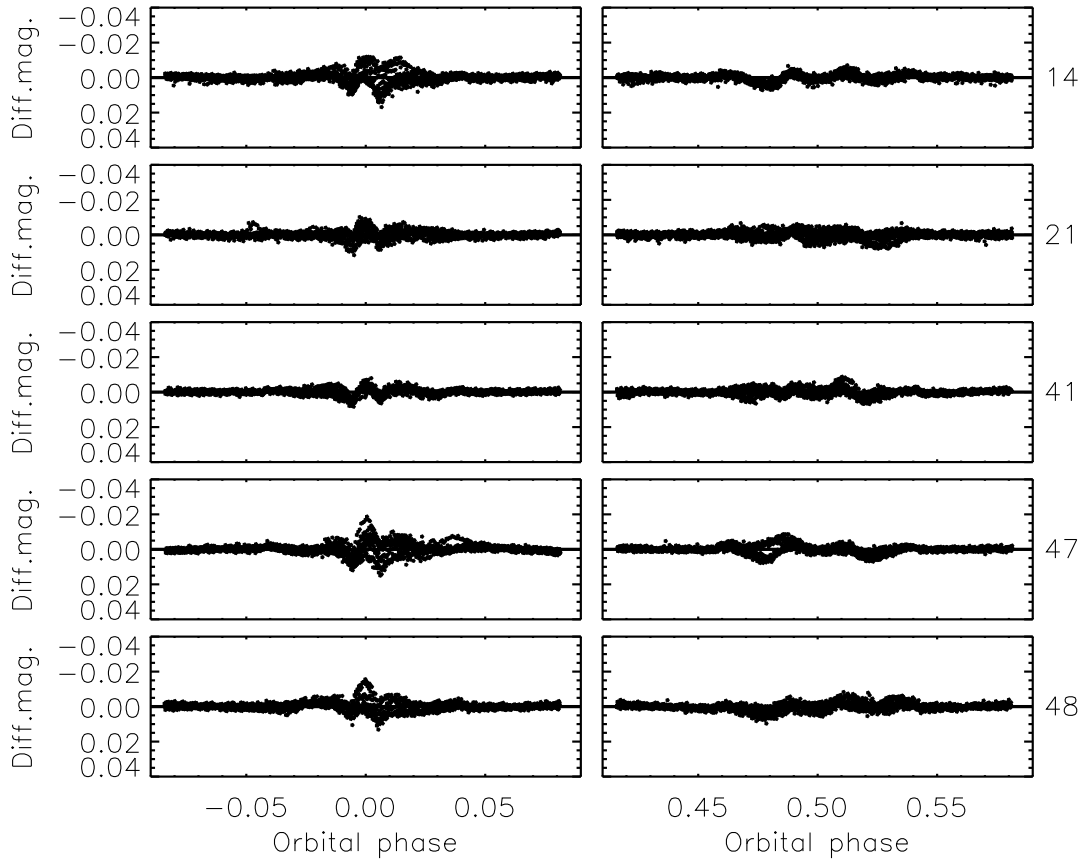} \\
\caption{\label{fig:resid2} Residuals of the best fits to individual \tess\ sectors, on the
same scale as Fig.~\ref{fig:resid1}. The sector numbers are shown on the right.} \end{figure}

The best fit is shown in Fig.~\ref{fig:phase} and has a large scatter during the eclipses. We attribute this to spots on the stellar surfaces which affect the surface brightness ratio of the system. The increased scatter during both eclipses is evidence that both components have starspots. A closer inspection of the residuals of the fit shows that their form changes between sectors (Fig.~\ref{fig:resid1}) due to evolution of the starspots.

From this we decided that separate fits to subsets of the \tess\ data are necessary. We therefore modelled the data from each sector individually with the same approach as above, except that $P$ was fixed at the value found from the fit to all data near eclipse. The residuals of these fits are shown in Fig.~\ref{fig:resid2}, which has been constructed in the same way as Fig.~\ref{fig:resid1} so the two figures may be easily compared. It is obvious that these individual fits give much lower residuals. This can be quantified by the scatter of the datapoints versus the best fit(s), which is 6.5~mmag for the overall fit and between 1.8 and 2.5~mmag for the individual fits. Nevertheless, some structure remains in the residuals in Fig.~\ref{fig:resid2} because the effects of starspots have not been completely removed.

The expectation from the fits to the data from the individual \tess\ sector was that the fractional radii would be reasonably consistent, as they are well determined by the contact points during the eclipses \cite{Russell12apj,Kopal59book}, but that the radiative parameters ($J$, $u_{\rm A}$, $u_{\rm B}$) would change significantly with time. Such assertions require the availability of errorbars, so we ran 1000 Monte Carlo simulations \cite{Me++04mn2,Me08mn} to measure the 1$\sigma$ uncertainties in the fitted parameters. We also experimented with further subdividing the data from each sector into two light curves, before and after the mid-sector pause for \tess\ to downlink data to Earth. The latter results are more informative so were adopted for the following analysis.

\begin{figure}[t] \centering \includegraphics[width=\textwidth]{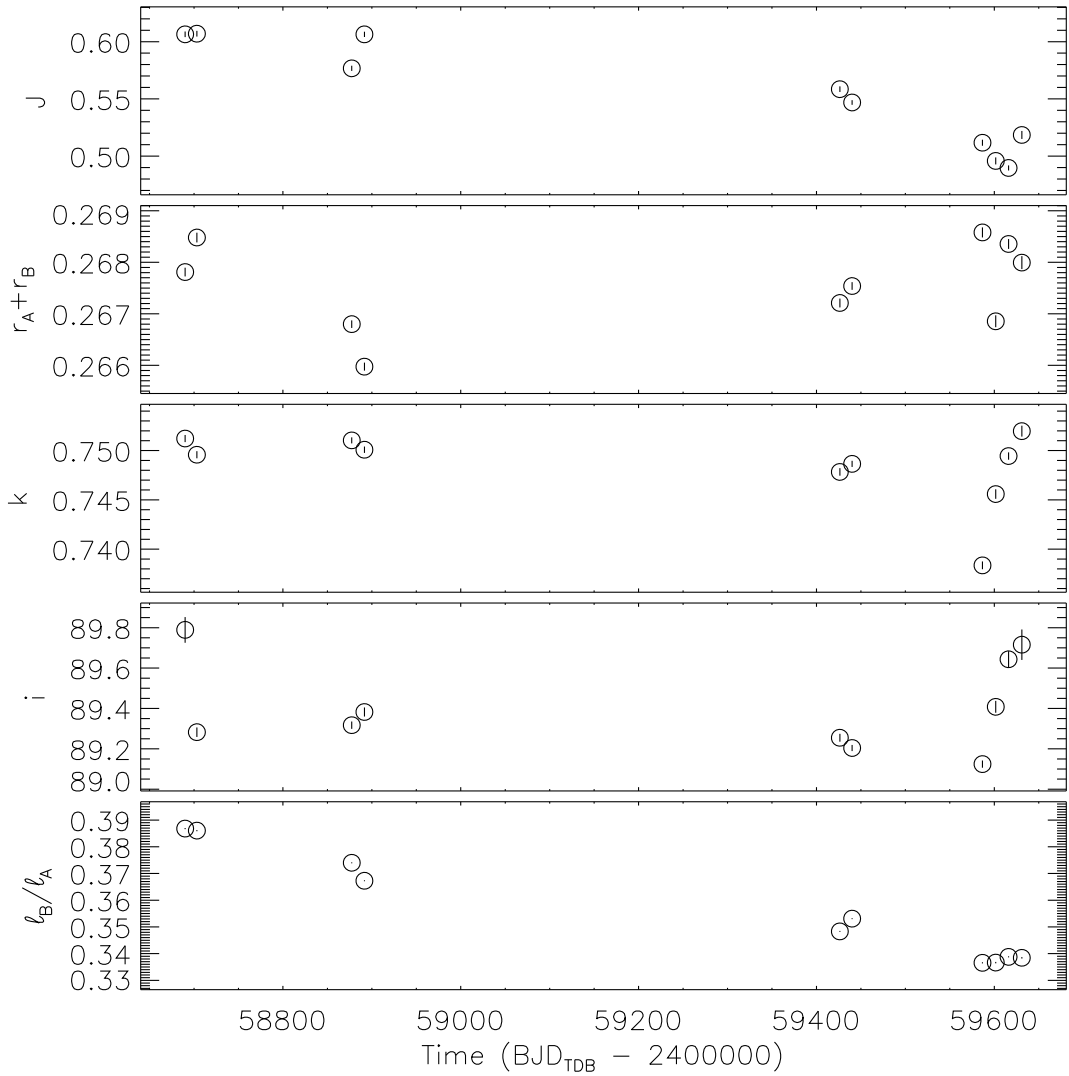} \\
\caption{\label{fig:sector} Values of important parameters measured from the eclipse
light curves of each half-sector (open circles). The errorbars come from 1000 Monte
Carlo simulations in each case and are all smaller than the point sizes. The points
are plotted versus the mean value of the timestamps on the x-axis.} \end{figure}

The results of this process are shown in Fig.\,\ref{fig:sector}, from which three conclusions can be reached. First, there is a clear trend of decreasing surface brightness and light ratio with time indicative of variation in the radiative parameters of one or both stars. Its timescale is a few years or more, but the current data are insufficient to infer this with any precision. Second, the wavelength-independent parameters ($r_{\rm A}+r_{\rm B}$, $k$, $i$) show no clear trend except for a possible variation in $k$ and $i$ for sectors 47 and 48. This agrees with prior expectations. Third, the errorbars are much smaller than the scatter of the values of individual parameters. This was also expected because they do not allow for the changing radiative parameters of the stars.

In the absence of an extensive study of the spot evolution properties of the system, we consider it reasonable to adopt the mean and standard deviation of the parameter values as the final values and errorbars. We use the standard deviation rather than the standard error to avoid underestimation of the uncertainties caused by the changing properties of the system. The adopted properties are given in Table~\ref{tab:jktebop}. The good news is that $r_{\rm A}$ and $r_{\rm B}$, the most important parameters here, are measured to 0.4\% precision. For completeness, the analysis based on the five light curves from individual sectors returned results in good agreement but with slightly smaller errorbars.

\begin{table} \centering
\caption{\em \label{tab:jktebop} Adopted parameters of \targ\ measured from the
\tess\ light curves using the {\sc jktebop} code. They represent the mean and
standard deviation of the values found from fitting the ten half-sector eclipse
light curves individually.}
\setlength{\tabcolsep}{20pt}
\begin{tabular}{lc}
{\em Parameter}                           &       {\em Value}                 \\[3pt]
Orbital inclination (\degr)               & $      89.41    \pm  0.22       $ \\
Sum of the fractional radii               & $       0.26756 \pm  0.00085    $ \\
Ratio of the radii                        & $       0.7484  \pm  0.0040     $ \\
Central surface brightness ratio          & $       0.552   \pm  0.046      $ \\
Third light                               &              0.0  (fixed)         \\
Linear LD coefficient for star A          & $       0.328   \pm  0.049      $ \\
Linear LD coefficient for star B          & $      -0.031   \pm  0.088      $ \\
Quadratic LD coefficient for star A       &              0.30 (fixed)         \\
Quadratic LD coefficient for star B       &              0.23 (fixed)         \\
Fractional radius of star~A               & $       0.15304 \pm  0.00067    $ \\
Fractional radius of star~B               & $       0.11453 \pm  0.00043    $ \\
Light ratio $\ell_{\rm B}/\ell_{\rm A}$   & $       0.357   \pm  0.020      $ \\[3pt]
\end{tabular}
\end{table}


\section*{Orbital ephemeris}

One drawback of modelling the \tess\ light curve in multiple short segments is that a precise orbital period does not result. We therefore sought to obtain a precise orbital ephemeris based on data covering a much longer time span. We made no attempt to be exhaustive, as we note that no change in orbital period is apparent in the many timings collected on the TIming DAtabase at Krakow (TIDAK\footnote{\texttt{https://www.as.up.krakow.pl/ephem/}}) for \targ\footnote{\texttt{https://www.as.up.krakow.pl/minicalc/UMAZZ.HTM}}. (see Ref.~\cite{Kreiner++01book}). The specific aim was to have an orbital ephemeris that is reliable over the time interval covering the recent \tess\ observations and the earlier spectroscopic studies in the mid-1990s that will be used below.

We began with the times of primary eclipse obtained from the ten half-sector eclipse light curves. The errorbars from the Monte Carlo analysis were significantly too small so we multiplied them by a factor of 20 to better account for their scatter versus a fitted linear ephemeris. To these we added published times of primary minimum obtained using CCDs or photoelectric photometers, plus the zeropoint of the orbital ephemeris given by Mallama \cite{Mallama80apjs}. These values were in all cases quoted in HJD, so we converted them to BJD. When a time system was given it was always UTC, so we assumed that all published timings were on the UTC system and converted them to TDB to match those from \tess, using routines from Eastman \etal\ \cite{Eastman++10pasp}.

\begin{figure}[t] \centering \includegraphics[width=\textwidth]{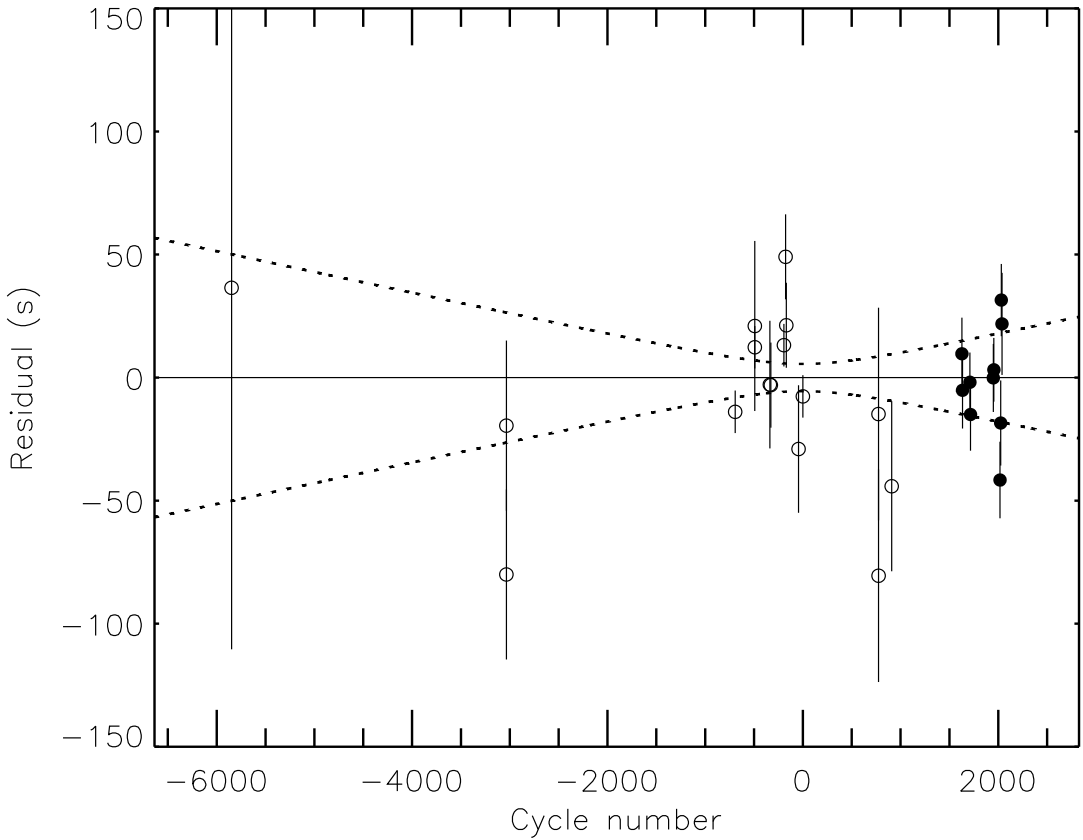} \\
\caption{\label{fig:tmin} Observed minus calculated ($O-C$) diagram of the
times of primary minimum versus the fitted linear ephemeris. Timings from the
\tess\ data are shown with filled circles. Timings from the literature are
shown with open circles. The dotted lines indicate the $1\sigma$ uncertainty
in the ephemeris determined from these data.} \end{figure}

The resulting orbital ephemeris is
\begin{equation}
\mbox{Min~I} = {\rm BJD}_{\rm TDB}~ 2454945.669574 (49) + 2.299260291 (39) E
\end{equation}
where $E$ is the cycle number since the reference time and the bracketed quantities indicate the uncertainties in the last digit of the preceding number. The reduced $\chi^2$ of the fitted ephemeris is $\chir = 1.75$ so the errorbars in the ephemeris above have been multiplied by $\sqrt{\chir}$ to account for this. This excess scatter in the eclipse timings beyond the quoted errorbars can be attributed to the spot activity shown by \targ. The times of minimum used in this analysis are given in Table~\ref{tab:tmin} and the residuals of the fit are shown in Fig.~\ref{fig:tmin}.

\begin{table} \centering
\caption{\em Times of published mid-eclipse for \targ\ and their residuals versus the fitted ephemeris.\label{tab:tmin}}
\setlength{\tabcolsep}{10pt}
\begin{tabular}{rcccl}
{\em Orbital} & {\em Eclipse time} & {\em Uncertainty } & {\em Residual } & {\em Reference} \\
{\em cycle}   & {\em (BJD$_{TDB}$)}    & {\em (d)}          & {\em (d)}       &              \\[3pt]
$-5848.0$ & 2441499.59581 & 0.00170 & $+$0.00042 & \cite{Mallama80apjs}         \\        
$-3035.0$ & 2447967.41436 & 0.00040 & $-$0.00023 & \cite{Hanzl91ibvs}           \\        
$-3035.0$ & 2447967.41366 & 0.00040 & $-$0.00093 & \cite{Hanzl91ibvs}           \\        
$ -693.0$ & 2453352.28203 & 0.00010 & $-$0.00016 & \cite{Kim+06ibvs}            \\        
$ -492.0$ & 2453814.43365 & 0.00010 & $+$0.00014 & \cite{Biro+07ibvs}           \\        
$ -492.0$ & 2453814.43375 & 0.00040 & $+$0.00024 & \cite{Hubscher++06ibvs}      \\        
$ -338.0$ & 2454168.51956 & 0.00030 & $-$0.00003 & \cite{Hubscher++09ibvs}      \\        
$ -328.0$ & 2454191.51216 & 0.00020 & $-$0.00004 & \cite{Hubscher++09ibvs}      \\        
$ -177.0$ & 2454538.70107 & 0.00020 & $+$0.00057 & \cite{Samolyk08javso}        \\        
$ -170.0$ & 2454554.79557 & 0.00020 & $+$0.00025 & \cite{Samolyk08javso}        \\        
$ -194.0$ & 2454499.61323 & 0.00010 & $+$0.00015 & \cite{Brat+08oejv}           \\        
$  -44.0$ & 2454844.50178 & 0.00030 & $-$0.00034 & \cite{Hubscher+10ibvs}       \\        
$    0.0$ & 2454945.66949 & 0.00010 & $-$0.00009 & \cite{Samolyk10javso}        \\        
$  774.0$ & 2456725.29687 & 0.00050 & $-$0.00017 & \cite{HubscherLehmann15ibvs} \\        
$  775.0$ & 2456727.59537 & 0.00050 & $-$0.00093 & \cite{HubscherLehmann15ibvs} \\        
$  909.0$ & 2457035.69667 & 0.00040 & $-$0.00051 & \cite{Hubscher15ibvs}        \\        
$ 1628.0$ & 2458688.86544 & 0.00017 & $+$0.00011 & \tess\ (this work)           \\        
$ 1634.0$ & 2458702.66083 & 0.00018 & $-$0.00006 & \tess\ (this work)           \\        
$ 1710.0$ & 2458877.40465 & 0.00014 & $-$0.00002 & \tess\ (this work)           \\        
$ 1716.0$ & 2458891.20006 & 0.00017 & $-$0.00017 & \tess\ (this work)           \\        
$ 1949.0$ & 2459426.92788 & 0.00016 & $-$0.00000 & \tess\ (this work)           \\        
$ 1955.0$ & 2459440.72348 & 0.00015 & $+$0.00004 & \tess\ (this work)           \\        
$ 2018.0$ & 2459585.57636 & 0.00018 & $-$0.00048 & \tess\ (this work)           \\        
$ 2025.0$ & 2459601.67145 & 0.00020 & $-$0.00021 & \tess\ (this work)           \\        
$ 2031.0$ & 2459615.46759 & 0.00017 & $+$0.00036 & \tess\ (this work)           \\        
$ 2038.0$ & 2459631.56230 & 0.00024 & $+$0.00025 & \tess\ (this work)           \\[3pt]   
\end{tabular}
\end{table}

\section*{Radial velocities}

Two sets of high-quality spectroscopic orbits for both components of \targ\ are available. Lacy \& Sabby \cite{LacySabby99ibvs} obtained 27 high-resolution spectra and measured RVs for both components in each. Imbert \cite{Imbert02aa} presented 33 RVs for star~A, of which 28 were from \textit{Coravel} \cite{Baranne++79va} and five from \textit{\'Elodie} \cite{Baranne+96aas}, and 16 RVs for star~B, of which 12 were from \textit{Coravel} and four from \textit{\'Elodie}. The \textit{Coravel} RVs have a greater scatter than those from Lacy \& Sabby, whereas the \textit{\'Elodie} RVs have a much lower scatter.

\begin{figure}[t] \centering \includegraphics[width=\textwidth]{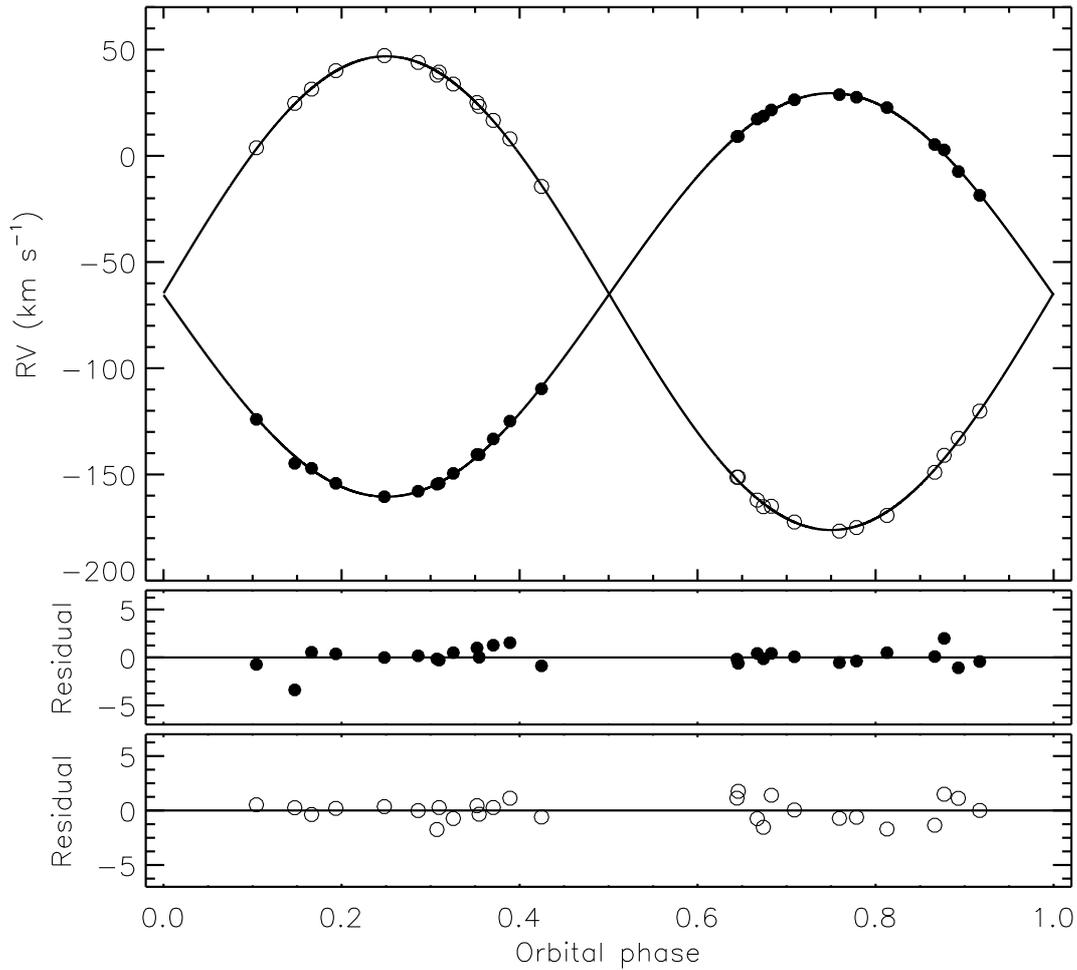} \\
\caption{\label{fig:lacy} RVs of \targ\ measured by Lacy \& Sabby \cite{LacySabby99ibvs}
(filled circles for star~A and open circles for star~B) compared to the best-fitting
spectroscopic orbits from {\sc jktebop} (solid curves). The residuals are given in
the lower panels separately for the two components.} \end{figure}

We reanalysed both sets of RVs independently, in order to ensure the published velocity amplitudes, $K_{\rm A}$ and $K_{\rm B}$, and uncertainties were reliable. In each case we modelled the RVs of both stars together but fitted for the individual systemtic velocities, $V_{\rm \gamma,A}$ and $V_{\rm \gamma,B}$. A circular orbit was assumed and the ephemeris was fixed at that found in the previous section. Informed by our work \cite{Me21obs5} on V505~Per we used 1000 Monte Carlo simulations to determine the errorbars for the fitted parameters.

The results for the Lacy \& Sabby \cite{LacySabby99ibvs} RVs are shown in Fig.~\ref{fig:lacy}. No data errors were given for the RVs so we weighted them equally for each star. Our results are in good agreement with those from Lacy \& Sabby \cite{LacySabby99ibvs}, and our errorbars are slightly smaller.

\begin{figure}[t] \centering \includegraphics[width=\textwidth]{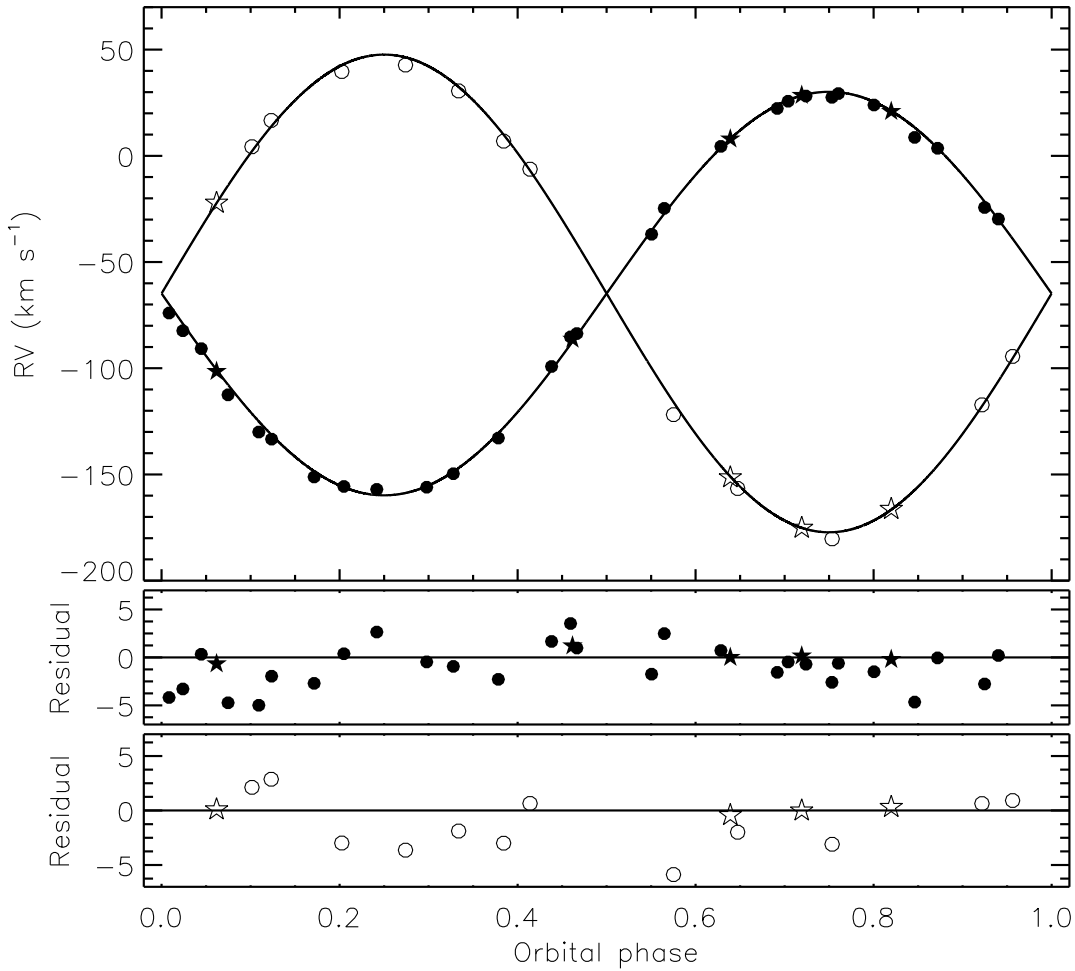} \\
\caption{\label{fig:imbert} RVs of \targ\ measured by Imbert \cite{Imbert02aa}
(filled symbols for star~A and open symbols for star~B) compared to the best-fitting
spectroscopic orbits from {\sc jktebop} (solid curves). The residuals are given in
the lower panels separately for the two components. The circles show RVs from
\textit{Coravel} and the stars show RVs from \textit{\'Elodie}. The axis ranges are
the same as those in Fig.~\ref{fig:lacy} so they may be compared easily.} \end{figure}

The fitted orbits for the Imbert \cite{Imbert02aa} RVs are shown in Fig.~\ref{fig:imbert}. We found that the uncertainties quoted for the \textit{\'Elodie} RVs for star~A were significantly smaller than the scatter of the RVs themselves so we doubled them for both stars. The RV uncertainties for each star were subsequently scaled to give $\chir = 1.0$. We found a slight disagreement in $K_{\rm B}$ for the Imbert RVs. Investigating this showed that the $K_{\rm B}$ is sensitive to the $T_0$ value used. We therefore recalculated all the orbits with $T_0$ fitted but $P$ fixed. This effect may be due to an inaccurate orbital ephemeris or to errors in the timestamps provided with the Imbert RVs. As the same problem was not found for the Lacy RVs, we suspect the latter.

Table~\ref{tab:orbits} presents the parameters of the spectroscopic orbits from the literature and found by ourselves. There is excellent consistency between different results. To obtain the final $K_{\rm A}$ and $K_{\rm B}$ we calculated the weighted mean of the individual values. The choice of whether to fix or fit $T_0$ has an effect on the final $K_{\rm A}$ and $K_{\rm B}$ values smaller than their uncertainties.

\begin{table} \centering
\caption{\em \label{tab:orbits} Spectroscopic orbits for \targ\ from the literature
and from the reanalysis of the RVs in the current work. All quantities are in\kms.}
\begin{tabular}{lrrrrrc}
{\em Source}  & $K_{\rm A}$~ & $K_{\rm B}$~ & ${\Vsys}$~ & ${\Vsys}_{\rm ,A}$~ & ${\Vsys}_{\rm ,B}$~ & $rms$ residual \\[7pt]

Lacy \& Sabby \cite{LacySabby99ibvs} &     95.1  &    111.8  &           &  $-$65.7  &  $-$64.8  \\
                                     & $\pm$0.3  & $\pm$0.3  &           & $\pm$0.2  & $\pm$0.2  \\
This work                            &     95.01 &    111.52 &           &  $-$65.50 &  $-$64.65 & 0.96, 0.96 \\
                                     & $\pm$0.23 & $\pm$0.24 &           & $\pm$0.18 & $\pm$0.18 \\[7pt]
Imbert \cite{Imbert02aa}             &     94.99 &    112.03 &  $-$65.22 &           &           & 1.73, 1.96 \\
                                     & $\pm$0.33 & $\pm$0.39 & $\pm$0.20 &           &           \\
This work                            &     94.89 &    111.75 &           &  $-$64.86 &  $-$65.36 & 2.00, 2.05 \\
                                     & $\pm$0.18 & $\pm$0.27 &           & $\pm$0.14 & $\pm$0.23 \\[7pt]
Final values                         &     94.94 &    111.62 \\
                                     & $\pm$0.18 & $\pm$0.40 \\
\end{tabular}
\end{table}



\section*{Chromospheric emission}

Magnetic fields in low-mass stars cause starspots \cite{Hale08apj,ThomasWeiss08book} and chromospheric emission in lines such as Ca~{\sc ii} H and K \cite{Wilson68apj,Noyes+84apj,Baliunas+95apj}. \targ\ shows evidence for the former phenomenon, and we had an opportunity to observe the latter. We obtained a single spectrum of \targ\ on the night of 2022/06/08 using the Intermediate Dispersion Spectrograph (IDS) at the Cassegrain focus of the Isaac Newton Telescope (INT). Thin cloud was present, which decreased the count rate of the observations. We used the 235~mm camera, H2400B grating, EEV10 CCD and a 1~arcsec slit in order to obtain a resolution of approximately 0.5~\AA, the maximum currently available with this spectrograph. A central wavelength setting of 4050\,\AA\ yielded a spectrum covering 373--438~nm at a reciprocal dispersion of 0.023~nm~px$^{-1}$. The data were reduced using a pipeline currently being written by the author, which performs bias subtraction, division by a flat-field from a tungsten lamp, aperture extraction, and wavelength calibration using copper-argon and copper-neon arc lamp spectra.

\begin{figure}[t] \centering \includegraphics[width=\textwidth]{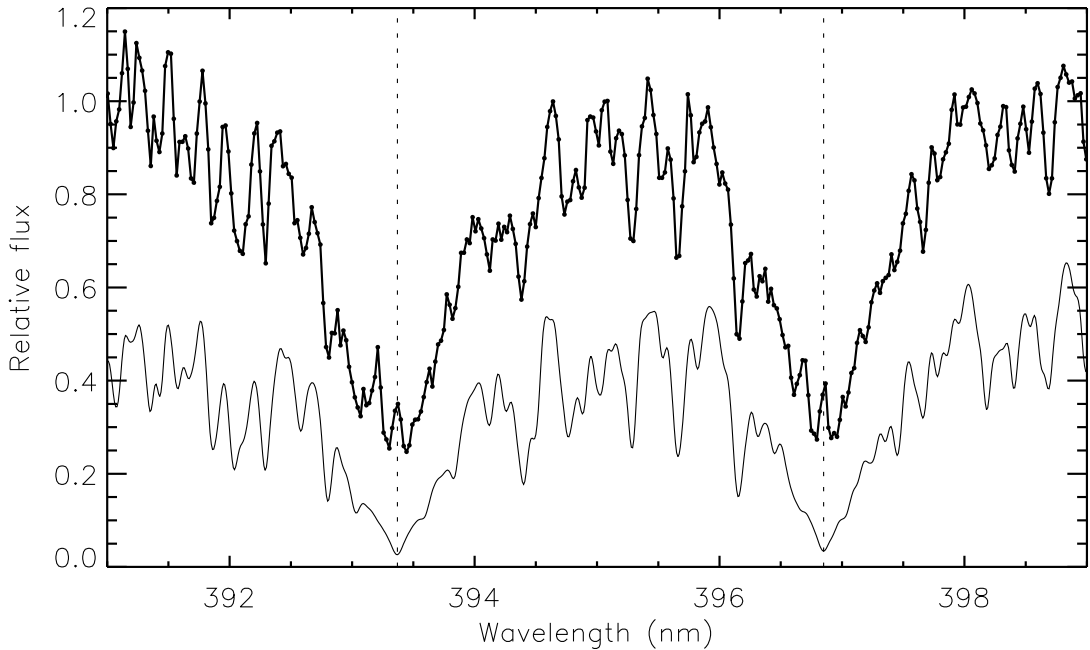} \\
\caption{\label{fig:cahk} Observed spectrum of \targ\ around the calcium H and K lines (thick upper line with
points) compared to a synthetic spectrum for a star with $\Teff = 6000$~K, $\logg = 4.5$ and solar metallicity
from the BT-Settl model atmospheres \cite{Allard+01apj,Allard++12rspta} (thin lower line). The H and K line central
wavelengths are shown with dotted lines. The spectrum of \targ\ has been shifted to zero velocity and normalised to
unit flux. The synthetic spectrum has been scaled so as to be clearly visible below the observed spectrum.} \end{figure}

Fig.~\ref{fig:cahk} shows the resulting spectrum in the region of the H and K lines compared to an analogous synthetic spectrum from the BT-Settl model atmospheres \cite{Allard+01apj,Allard++12rspta}. Emission in the cores of the calcium lines is obvious and confirms that the system shows chromospheric activity. The spectrum was taken at orbital phase 0.9037 where the RV difference between the stars was 117\kms\ (approximately 3.5 pixels). The emission lines from the two stars overlap so it is not possible to assess the emission strengths from the two stars individually.

\section*{Physical properties of \targ}

The physical properties of the system were calculated using the $r_{\rm A}$, $r_{\rm B}$ and $i$ from Table~\ref{tab:jktebop}, the orbital period determined above, and the final $K_{\rm A}$ and $K_{\rm B}$ values from Table~\ref{tab:orbits}. We used standard formulae \cite{Hilditch01book} and the reference solar values from the IAU \cite{Prsa+16aj}, as implemented in the {\sc jktabsdim} code \cite{Me++05aa}. The results are given in Table~\ref{tab:absdim} and show good news: the masses are measured to precisions of 0.8\% and 0.5\%, and the radii to 0.5\% and 0.4\%. The main limitation is the lower precision of $K_{\rm B}$, which is in turn due to the minor disagreement between the two sources of published RVs.

A comparison with the properties measured by Lacy \& Sabby \cite{LacySabby99ibvs} shows excellent agreement for the masses (as expected) but not for the radii: we find values significantly lower than the $1.51 \pm 0.02$ and $1.16 \pm 0.01$\Rsun\ given by Lacy \& Sabby \cite{LacySabby99ibvs}. We expect our own radius measurements to be more reliable as they are based on light curves of much better quality and greater quantity than in previous studies.

The effective temperatures (\Teff s) of the stars were determined by Lacy \& Sabby \cite{LacySabby99ibvs} based on the intrinsic $(b-y)_0$ colour indices of the stars, which come from the combined colour indices, the light ratios in the $b$ and $y$ light curves from Clement \etal\ \cite{Clement+97aas}, and the intrinsic flux calibration by Popper \cite{Popper80araa}. This approach is at first impression quite outdated, but can be checked using external information. First, we used the surface flux ratio in Table~\ref{tab:jktebop} to determine an approximate \Teff\ ratio of $0.862 \pm 0.018$ which gives $\TeffB = 5138 \pm 107$~K for a $\TeffA$ of 5960~K, in acceptable agreement with the $\TeffB$ from Lacy \& Sabby \cite{LacySabby99ibvs}. This point is only weak evidence because the variation in radiative parameters seen in the \tess\ light curves means our measurement of the surface brightness ratio may not be representative of its average value. Second, we determined the distance to the system using the method of Southworth \etal\ \cite{Me++05aa} which relies on the surface brightness versus \Teff\ calibrations of Kervella \etal\ \cite{Kervella+04aa}. The $K$-band measurement gives a distance of $180.8 \pm 1.9$~pc assuming an interstellar extinction of $\EBV = 0.00 \pm 0.01$; a larger interstellar extinction can be ruled out by requiring the distances found in the $BVJHK$ bands to be consistent. The \gaia\ EDR3 parallax \cite{Gaia21aa} gives a distance of $180.20 \pm 0.45$~pc by simple inversion. Based on this, we accepted the \Teff\ values from Lacy \& Sabby \cite{LacySabby99ibvs} as reliable.

\begin{table} \centering
\caption{\em Physical properties of \targ\ defined using the nominal solar units given by IAU
2015 Resolution B3 (Ref.\ \cite{Prsa+16aj}). \label{tab:absdim}}
\begin{tabular}{lr@{\,$\pm$\,}lr@{\,$\pm$\,}l}
{\em Parameter}        & \multicolumn{2}{c}{\em Star A} & \multicolumn{2}{c}{\em Star B}    \\[3pt]
Mass ratio                                  & \multicolumn{4}{c}{$0.8505 \pm 0.0034$}       \\
Semimajor axis of relative orbit (\Rsunnom) & \multicolumn{4}{c}{$9.388 \pm 0.020$}         \\
Mass (\Msunnom)                             &  1.1348 & 0.0087      &  0.9652 & 0.0051      \\
Radius (\Rsunnom)                           &  1.4367 & 0.0070      &  1.0752 & 0.0046      \\
Surface gravity ($\log$[cgs])               &  4.1782 & 0.0041      &  4.3597 & 0.0034      \\
Density ($\!\!$\rhosun)                     &  0.3826 & 0.0051      &  0.7765 & 0.0089      \\
Synchronous rotational velocity ($\!\!$\kms)&  31.61  & 0.15        &  23.66  & 0.10        \\
Effective temperature (K)                   &   5960  & 70          &   5270  & 90          \\
Luminosity $\log(L/\Lsunnom)$               &   0.370 & 0.021       &$-$0.095 & 0.030       \\
$M_{\rm bol}$ (mag)                         &   3.814 & 0.052       &   4.978 & 0.075       \\
Distance (pc)                               & \multicolumn{4}{c}{$180.8 \pm 1.9$}           \\[3pt]
\end{tabular}
\end{table}




\section*{Comparison with theoretical models}

\begin{figure}[t] \centering \includegraphics[width=\textwidth]{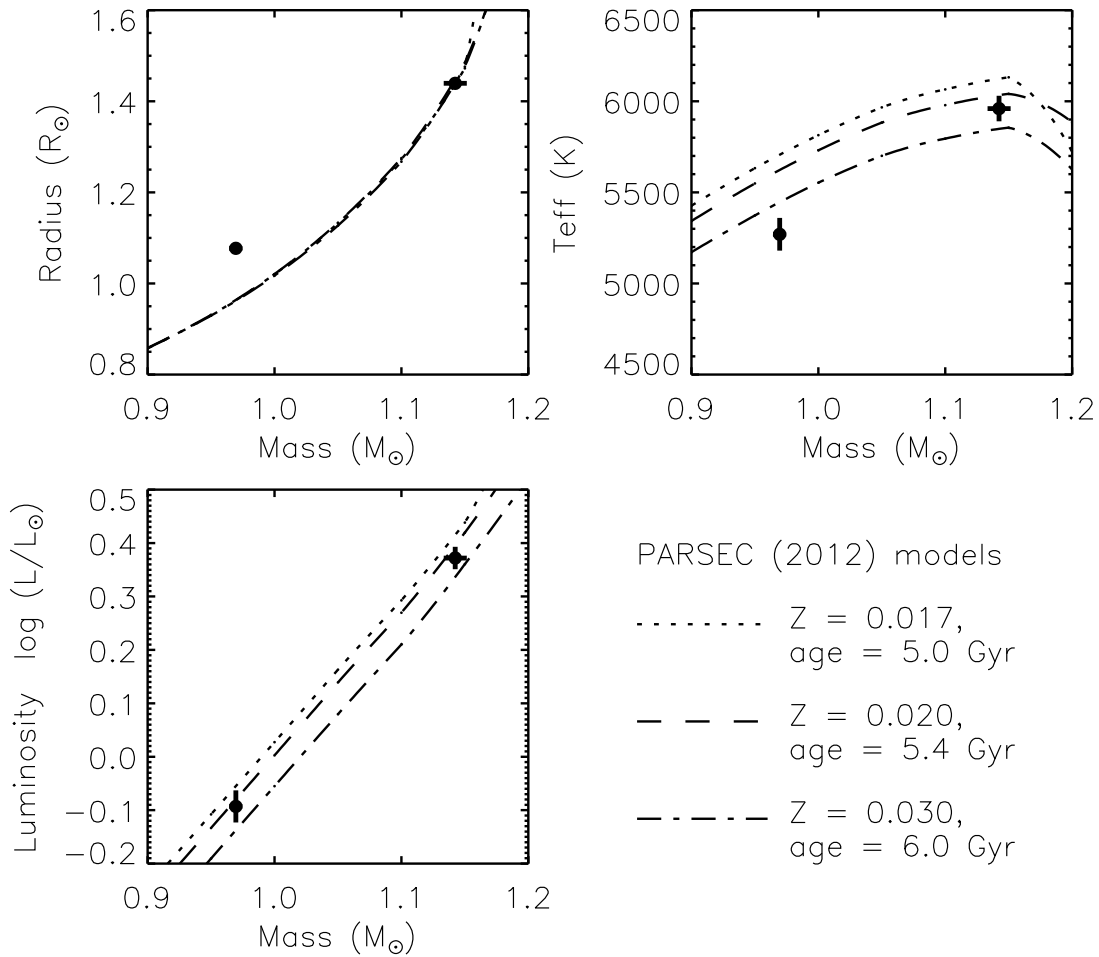} \\
\caption{\label{fig:model} Comparison between theoretical predictions from the
{\sc parsec} models \cite{Bressan+12mn} and the measured physical properties of
\targ\ for mass versus, radius, \Teff\ and luminosity. The ages and metal abundances
of the chosen theoretical predictions are given in the key on the lower right.} \end{figure}

The precise measurement of the properties of the stars, in particular for the slightly-evolved star~A, means a comparison with theoretical evolutionary models could be informative. We did this, using tabulations from the {\sc parsec} models \cite{Bressan+12mn}, via the mass--radius and mass--\Teff\ diagrams \cite{MeClausen07aa,Me22obs4} (Fig.~\ref{fig:model}). We did not consider any constraints on metallicity as there are no precise determinations available.

The mass, radius and \Teff\ of star~A can be matched by a metallicity in the region of $Z=0.02$ or slightly higher, with an age in the region of 5.5~Gyr depending on the adopted $Z$. Theoretical models with $Z=0.017$ predict a significantly higher \Teff\ than observed, and models with $Z=0.030$ predict the opposite. We infer that the system has an approximately solar metallicity and that star~A does not contradict stellar theory.

The less massive star~B, however, causes a problem for the theoretical models. It is not possible to match its properties for any combination of age and metallicity available in the {\sc parsec} tabulations. It is much too large and cool to match any theoretical predictions simultaneously with star~A (see Figure), and the closest we can get is for $Z=0.06$ and an age of 8.5~Gyr, where its radius is well matched but it is still slightly too hot for the models. By comparison to its predicted properties for the same age(s) and metal abundance(s) as star~A, it is 0.12\Rsun\ (11\%) larger and 330~K (6\%) cooler, but these effects cancel to give it the expected luminosity.

A \emph{radius discrepancy} is known to exist for late-type dwarfs in the sense that they have larger radii and lower \Teff s than predicted by stellar theory \cite{Spada+13apj,Torres13an,MorrellNaylor19mn,Hoxie73aa,Lacy77apjs}. This has been attributed to the inhibition of convection \cite{Lopez07apj,Feiden16aa} due to magnetic fields and/or starspots. There is evidence that stars in short-period binaries are affected in a qualitatively different way \cite{Lopez07apj} because tidal effects cause them to rotate more quickly and thus show stronger activity. However, the radius discrepancy has also been found in longer-period binaries \cite{Irwin+11apj} and single stars \cite{Spada+13apj,MorrellNaylor19mn}, and there are also examples of stars in short-period binaries that do \emph{not} show the discepancy \cite{Blake+08apj,Feiden++11apj,Maxted+22mn}. Recent analyses have found a large scatter in the size of the radius discrepancy between and even within binary systems \cite{Kraus+17apj,Parsons+18mn}. Under the assumption that the age and metallicity of the \targ\ system can be inferred from the properties of star~A, the poor agreement between theoretical models and the measured properties of star~B make it an excellent candidate for a star showing the radius discrepancy. \targ~B is another example of a star in the region of 1\Msun\ showing the radius discrepancy.

An updated version of the {\sc parsec} model grid, denoted 1.2S, has been computed \cite{Chen+14mn} with a revised temperature vs.\ optical depth prescription based on sophisticated model atmospheres of low-mass stars \cite{Allard+01apj,Allard++12rspta}. These provide a better match to the observed properties of low-mass stars. However, they are not useful in the current situation as the modifications cover only masses of 0.7\Msun\ and lower.

\section*{Summary and conclusions}

\targ\ is a solar-type dEB which shows total eclipses with an orbital period of 2.299~d and evolving starspot activity. We have determined its physical properties based on two high-quality sets of RVs available from the literature and light curves from five sectors of observations with the \tess\ satellite. The two RV datasets agree with each other very well for the primary star but not quite so well for the secondary star, but do allow mass determinations to 0.8\% and 0.6\%, respectively.

The light curve of \targ\ varies during and between \tess\ sectors due to starspot evolution, which manifests itself as a slow variation with time (as spots rotate in and out of view), emission in the calcium H and K lines, and changes in the depths of the eclipses. By modelling each \tess\ half-sector separately, we have found slow variations in the light and surface brightness ratios between the stars indicative of spot-induced changes in their radiative properties. The geometric parameters ($r_{\rm A}$, $r_{\rm B}$ and $i$) show much smaller changes because they are derived primarily from the contact points during the total and annular eclipses. We were therefore able to measure the radii of the stars to 0.5\% precision. Unfortunately, no further observations with \tess\ are scheduled for this system. It would be interesting to identify a similar dEB in either the \tess\ continuous viewing zones or that has been observed using the \kepler\ satellite, for which the spot evolution could be tracked in the same way over a longer and/or better-sampled time period.

The properties of star~A match predictions from the {\sc parsec} models for an age of approximately 5.5~Gyr and a slightly super-solar metal abundance around $Z=0.02$. Star~B does not match these or any other predictions, having too large a radius and too low a \Teff\ to agree with the theoretical models. Its luminosity is consistent with the models because the radius and \Teff\ offsets cancel. Star~B therefore is an excellent example of the \emph{radius discrepancy} well-known to affect low-mass eclipsing binaries, and at 0.97\Msun\ is one of the most massive stars to clearly show this effect.

An alternative interpretation is that \targ\ is younger than we infer from matching star~A to the predictions of theoretical models, and that \emph{both} stars show the radius discrepancy. This could be investigated using dEBs with external constraints on their ages (and chemical compositions), e.g.\ via membership of an open cluster \cite{Me++04mn,Brogaard+11aa,Brogaard+12aa}.


\section*{Acknowledgements}

We thank Drs.\ Guillermo Torres, Patricia Lampens and Pierre Maxted for comments on draft versions of this work.
This paper includes data collected by the \tess\ mission and obtained from the MAST data archive at the Space Telescope Science Institute (STScI). Funding for the \tess\ mission is provided by the NASA's Science Mission Directorate. STScI is operated by the Association of Universities for Research in Astronomy, Inc., under NASA contract NAS 5–26555.
The following resources were used in the course of this work: the NASA Astrophysics Data System; the SIMBAD database operated at CDS, Strasbourg, France; and the ar$\chi$iv scientific paper preprint service operated by Cornell University.


\bibliographystyle{obsmaga}

\end{document}